# Evaluation methods and empirical research on coastal environmental performance for Chinese harbor cities

Yi Zheng

*Business School, Shanghai Jian Qiao University, Shanghai 201306，China.*
*Email: yzheng@shou.edu.cn*

**Abstract**

For controlling pollution of the marine environment while developing coastal economy, the coastal environmental performance was proposed and measured in static and dynamic methods combined with DEA and efficiency theory in this paper. With the two methods, 16 harbor cities were evaluated. The results showed the index designed in this paper can better reflect the effect to the marine environment for economy of the coastal cities.

## 1. Introduction

In recent years, more than 80% national and regional planning focus of marine economy and sea development, more than 90% in the coastal city proposed the strategy of marine development, marine has become an important driving force for China's economic and social development in the new period. But with the rapid development of coastal economy, especially as the economic center, the rapid expansion of the harbor cities, the impact on the marine ecological environment becomes more and more significant. So it is an important guarantee for the sustainable development of economy in coastal region to maintain and improve the marine environment. This requires that the development of marine economy in our country should be properly adjusted. Evaluation for marine environmental performance can find the existing problem of marine economy in the coastal areas, so as to providing some reference for scientific decision-making and guide the coastal local economy to a healthy way.

In recent study on coordination of economy and environment, the concept of eco-efficiency has received increasing public attention and plays an important role in the business community and the governments. In ecological economics literature, eco-efficiency is commonly defined as a ratio of value added to environmental damage added where larger values of the ratio are considered as better environmental performance. Although a point in the critique is that even if the relative level of environmental pressure is low relative to economic output, the absolute environmental pressure can still exceed the carrying capacity of the ecosystem, this does not render the concept of eco-efficiency useless. Measurement of eco-efficiency is critically important for at least two reasons: (1) Improvement of eco-efficiency is often the most cost-effective way of reducing environmental pressures. Even if efficiency improvements as such may not suffice for achieving a sustainable level of environmental pressure, it makes economic sense to exploit these options as much as possible. (2) Policies targeted at efficiency improvements tend to be easier to adopt than policies that restrict the level of economic activity. Thus, measurement of eco-efficiency is of critical importance even if efficiency improvements as such prove insufficient.

For predictive models to provide reliable guidance in decision making processes, they are often required to be accurate and robust to distribution shifts (Zheng et al., 2022). The greatest challenge in constructing an eco-efficiency index and composite environmental performance indices in general is to aggregate various environmental pressures into a single environmental damage index. One of existing approaches toward environmental performance indicators is the so-called LIFE CYCLE ANALYSIS AND ASSESSMENT (LCA ), which is a tool for studying the impacts of a given product over all stages of its life cycle (resource extraction, energy use, production, distribution, use and ultimate disposal), However, even when such a methodology eventually becomes well understood and effective, it does not tell us how to integrate the impacts of the various products into one or a few plant-level indicators, which is the goal we are pursuing（Tyteca, 1996).

The other commonly environmental performance measures are in the scope of the theory of production efficiency. One of the earliest studies toward incorporating environmental preoccupation into production efficiency was that by Pittman (1983) who generalized previous work by Caves et al. (1982a,b) for coping with pollution, taken as an undesirable output. As stated by Pittman (1983), ''they are not estimated at the plant level'' but rather come from estimates at the state level, and may therefore not reflect the actual conditions encountered. More recently, Fare et al. (1993), used a parametric specification of a distance function as defined by Shephard (1970) and obtained the required specification through the solution of a linear programming problem where the shadow prices of the ''bads'' (the undesirable outputs) were imposed to be non-positive. In the other way, the non-parametric approach adopted by Fare et al. (1989) started from Farrell's (1957) measure of technical efficiency. Seiford and Thrall (1990) reviewed the various advantages of non-parametric approaches (including DEA) over parametric approaches. Among these advantages are the robustness of the linear programming methods used to solve DEA problems and the new insights and additional information it provides with respect to conventional econometric methods. Charnes et al. (1985) also noted that a variable that is neither an economic resource nor a product, but is an attribute of the environment or of the production process, can be included easily in a DEA-based production model. In addition, Kortelainen (2008) promoted environmental performance measures from static analysis to dynamic analysis with the combination of the malmquist index and the data envelopment analysis from the perspective of ecological efficiency.

Although various works for eco-efficiency measurement have been presented in the literature, study on marine environmental performance based on the concept of ecological efficiency and using data envelopment analysis method is almost not found. So based the research achievements of the eco-efficiency and environmental performance, this paper put forward the concept of coastal environmental performance, which measures the ocean environmental impact by economic development of coastal region. Then with the DEA method the concrete models of static and dynamic analysis for coastal environmental performance were given. At last, the evaluation on marine environmental performance was calculated for the 16 harbor city of the most economically developed coastal areas of China. The empirical studies show the evaluation method is effective.

## 2. Evaluation for coastal environmental performance

In this article we propose coastal environmental performance of production concerns the capability to produce goods and services in coastal area while causing minimal ocean environmental degradation. Here, the environmental damage of economic activities is defined as the damages to the coastal marine environment. Aggregation of environmental pressures into a single environmental damage index is a major challenge of environmental performance measurement. The data envelopment analysis (DEA) method can be adapted for this purpose. (Kuosmanen and Kortelainen, 2005).

**2.1　The static analysis for coastal environmental performance**

Under evaluation relative to a sample of N comparable units, let $V_n$ denote the economic value added and $Z_n$ environmental pressures of unit n (n = 1, . . . , N). For transparency, the capital symbols Vn, Zn refer to observed data of n, whereas arbitrary (theoretical) values of value added and environmental pressures are denoted by lower case v and z. So the eco-efficiency of unit n is expressed as EEn = Vn / D (Zn ), where D is the damage function that aggregates the M environmental pressures into a single environmental damage score.

In contrast to economic output $V_n$, environmental pressures do not typically have prices or other unambiguous values. This presents a challenge for constructing the environmental damage index $D(\mathbf{Z}_n)$. To build up an encompassing "total" eco-efficiency ratio, a natural approach is to use a linear approximation of $D$ and use a weighted sum of the various environmental pressures, that is, $D(\mathbf{z}) = w_1 z_1 + w_2 z_2 + \ldots + w_M z_M$, where wm (m = 1, . . . , M) represents the weight accorded to environmental pressure m. Formally, a eco-efficiency score of activity k can be calculated with the following equation.

$$EE_k = \max_w \frac{V_k}{\sum_{m=1}^{M} w_m Z_{km}}$$

$$\text{s.t.} \quad \frac{V_n}{\sum_{m=1}^{M} w_m Z_{nm}} \leq 1, \quad \forall n = 1, \Lambda, N \qquad (1)$$

$$w_m \geq 0, \quad \forall m = 1, \Lambda, M$$

Because it is a fractional linear programming problem involving a nonlinear objective function and nonlinear constraints, the problem would be linearized by solving the reciprocal problem

$$(EE_k)^{-1} = \min_w \sum_{m=1}^{M} w_m \frac{Z_{km}}{V_k}$$

$$\text{s.t.} \quad \sum_{m=1}^{M} w_m \frac{Z_{nm}}{V_n} \geq 1, \quad \forall n = 1, \Lambda, N \qquad (2)$$

$$w_m \geq 0, \quad \forall m = 1, \Lambda, M$$

This problem can be solved by standard linear programming algorithms. The relative eco-efficiency score is obtained by taking the inverse of the optimal solution to Eq. (2). In evaluation for coastal cities, the relative eco-efficiency is thought as static coastal environmental performance. An important property of the benefit of the doubt weighting scheme is that it does not demand any prior information concerning the weights of the different environmental pressures, the only constraint for weights in Eq. (2) is no negativity. By DEA theory, the eco-efficiency score given by Eq. (2) may be decomposed into pure technical efficiency and scale efficiency. The former represents an improvement of efficiency change induced by technology; the later represents the efficiency change produced by the different scale.

**2.2　the dynamic analysis for coastal environmental performance**

For the purpose of the dynamic analysis for coastal environmental performance, in this section, we present an environmental performance index by utilizing the Malmquist productivity index. Firstly, we need some additional notation. Let $EE_k(Z_s, V_s, t)$ denote the relative eco-efficiency measure of production unit k observed in period s, measured relative to the frontier of period t, calculated as follows:

$$[EE_k(Z^s, V^s, t)]^{-1} = \min_w \sum_{m=1}^{M} w_m \frac{Z_{km}(s)}{V_k(s)}$$

$$\text{s.t.} \quad \sum_{m=1}^{M} w_m \frac{Z_{nm}(t)}{V_n(t)} \geq 1, \quad \forall n = 1, \Lambda, N \quad (3)$$

$$w_m \geq 0, \quad \forall m = 1, \Lambda, M$$

where indices s, t in parentheses refer to the period of observation. To measure the change in the environmental performance of unit k from period t−1 to t, we can take the frontier of period t as the benchmark and quantify environmental performance change by the ratio of relative eco-efficiency scores based on adjacent observations. Formally

$$EPI_k(t) = \frac{EE_k(Z^t, V^t, t)}{EE_k(Z^{t-1}, V^{t-1}, t)} \quad (4)$$

where $EPI_k(t)$ means the environmental performance index of unit k and t in parentheses is the period of reference technology. Since we have no reason to prefer period t or t−1 as a benchmark for the index $EPI_k$, the geometric average in the two periods was given:

$$EPI_k(t-1, t) = [EPI_k(t-1) \times EPI_k(t)]^{1/2} = \left[ \frac{EE_k(Z^t, V^t, t-1)}{EE_k(Z^{t-1}, V^{t-1}, t-1)} \times \frac{EE_k(Z^t, V^t, t)}{EE_k(Z^{t-1}, V^{t-1}, t)} \right]^{1/2} \quad (5)$$

This environmental performance index (EPI) is technically analogous to the input-oriented Malmquist productivity index presented in the productive efficiency literature. The values greater than one indicate improvement of environmental performance over time, while values less than one indicate deterioration. It is unit invariant, i.e. it does not depend on the measurement units of model variables.

Following Nishimizu and Page (1982) and Färe et al. (1994b), we can decompose the overall coastal environmental performance change (Eq.(5)) into two subcomponents as follows:

$$EPI_k(t-1, t) = \frac{EE_k(Z^t, V^t, t)}{EE_k(Z^{t-1}, V^{t-1}, t-1)} \times \left[ \frac{EE_k(Z^{t-1}, V^{t-1}, t-1)}{EE_k(Z^{t-1}, V^{t-1}, t)} \times \frac{EE_k(Z^t, V^t, t-1)}{EE_k(Z^{t-1}, V^{t-1}, t)} \right]^{1/2} \quad (6)$$

In Eq.(6), the first ratio can be interpreted as a measure of efficiency change, as it reveals how production unit's environmental performance has changed relative to the benchmarks. A value greater (less) than one indicates that the unit has caught up (fallen behind) its benchmark in period t as compared to the benchmark in t−1, i.e. it has moved towards (away from) the efficient frontier. So it is representing the relative eco-efficiency change. The other subcomponent in Eq.(6) indicates a change in the environmental technology. It measures shifts in the eco-efficiency frontier or the best possible performance in period t as compared to period t−1. The value of the environmental technical change shows whether the best practice technology is improving, stagnant or deteriorating. Again, if its value is greater than one, it indicates that the environmental performance of the most eco-efficient units has improved. In practice, this kind of improvement in the environmental performance can be a consequence of some technological innovation or regulatory changes, for instance. Importantly, this subcomponent can reveal important information that cannot be obtained from the static eco-efficiency analysis.

## 3. An empirical study on Chinese harbor cities

### 3.1 Index and data

Economic development in China, especially in the opening harbor city, has great effects on the marine environment. Therefore, to verify the effectiveness of the index and the evaluation method of coastal environment performance, Chinese 16 representative harbor cities were evaluated by the above method. Because the influence of environment here is defined as the damages to the marine environment, so according to the availability of data, the emissions of a direct impact on the marine environment were chosen as environmental pressures index. They are amount of ammonia nitrogen in industrial waste water, amount of COD in industrial waste water and volume of industrial solid wastes generated. The increased value of economic activity in the measurement model is used of Gross Regional Product. The data in this paper was drawn from China statistical yearbook of environment and China statistical yearbook of regional economy of the year from 2011 to 2013, and the calculation with model (3), (5) is made use of the DEA-Solver Pro 9 Software

Tab.1  The statistical description of index for 16 harbor cities from 2010 to 2012

| Time period | Statistic | Amount of Ammonia Nitrigen in Industrial Waste Water (ton) | Amount of COD in Industrial Waste Water (ton) | Volume of Industrial Solid Wastes Generated (10000 tons) | Gross Regional Product (100 million yuan) |
|---|---|---|---|---|---|
| 2010 | Max | 3197 | 26249 | 2448.4 | 17165.98 |
|  | Min | 269 | 2317 | 101.1 | 401.41 |
|  | Average | 951.75 | 12655.5 | 789.9688 | 4423.383 |
|  | SD | 900.0452 | 7195.617 | 767.5555 | 4277.658 |
| 2011 | Max | 3253 | 27357 | 2442.2 | 19195.69 |
|  | Min | 112 | 4313 | 95.77 | 496.58 |
|  | Average | 1156.938 | 13679.81 | 877.385 | 5174.569 |
|  | SD | 890.7748 | 7103.952 | 806.078 | 4903.085 |
| 2012 | Max | 3295 | 26533 | 2565.38 | 20181.72 |
|  | Min | 129 | 4080 | 105.23 | 630.09 |
|  | Average | 1120.938 | 13683.75 | 881.0094 | 4925.67 |
|  | SD | 863.3378 | 7044.606 | 817.6447 | 5074.165 |

Data sources: 2011- 2013 China statistical yearbook of environment and China statistical yearbook of regional economy.

Tab.2  Sea Area with Water Quality below Grade 1 in near shore from 2010 to 2012

sq. km

| Sea Area | Sea Area with Water Quality at Grade 2 | | | Sea Area with Water Quality at Grade 3 | | | Sea Area with Water Quality at Grade 4 | | | Sea Area with Water Quality below Grade 4 | | |
|---|---|---|---|---|---|---|---|---|---|---|---|---|
|  | 2010 | 2011 | 2012 | 2010 | 2011 | 2012 | 2010 | 2011 | 2012 | 2010 | 2011 | 2012 |

| | | | | | | | | | | | |
|---|---|---|---|---|---|---|---|---|---|---|---|
| BohaiSea | 14600 | 14690 | 12330 | 13600 | 8950 | 11040 | 4600 | 3790 | 4690 | 2700 | 4210 | 13080 |
| YellowSea | 27700 | 13780 | 12890 | 8600 | 7170 | 3450 | 4800 | 4240 | 7540 | 200 | 9540 | 16530 |
| EastChinaSea | 22600 | 15430 | 12800 | 16300 | 10820 | 7540 | 6600 | 9150 | 8820 | 35800 | 27270 | 33970 |
| SouthChinaSea | 25700 | 3940 | 8890 | 6100 | 7370 | 8000 | 0 | 1160 | 3650 | 2100 | 2780 | 4300 |

Data sources: 2011- 2013 China statistical yearbook of environment.

### 3.2 The static and dynamic analysis for coastal environmental performance of 16 harbor cities

In this section, based the 2012 static data about Tianjin, Qinhuangdao, Dalian, Lianyungang, Qingdao, Yantai, Shanghai, Ningbo, Wenzhou, Fuzhou, Xiamen, Shenzhen, Zhuhai, Shantou, Zhanjiang and Beihai (in Tab.1), the static coastal environmental performance was calculated with Eq.(3). The results were listed in column B of Tab.3. Then it was decomposed into pure technical efficiency and scale efficiency according to DEA theory, and the results were listed in column C and D in Tab.3. Based these data, the Dynamic environmental performance from 2010 to 2012 was calculated with Eq.(5), and then the result was decomposed into dynamic efficiency change and dynamic technical change with Eq.(6) and was listed in column E and F of Tab.3。

Tab3. Coastal environmental performance of 16 harbor cities and pollution comparing in the sea area near these cities

| City (A) | Static environmental performance in 2012 (B) | Static technical efficiency in 2012 (C) | Static scale efficiency in 2012 (D) | Dynamic efficiency change from 2010 to 2012 (E) | Dynamic technical change from 2010 to 2012 (F) | Dynamic environmental performance from 2010 to 2012 (G) | Sea Area near the city (H) | Average static environmental performance in 2012 (I) | Increased sea Area with Water Quality below Grade 1 from 2011 to 2012 (J) | Average dynamic environmental performance from 2010 to 2012 (K) | Increased sea Area with Water Quality below Grade 1 from 2010 to 2012 (L) |
|---|---|---|---|---|---|---|---|---|---|---|---|
| Tianjin | 3% | 15% | 17% | 0.14 | 0.43 | 0.06 | | | | | |
| Qinhuangdao | 11% | 35% | 32% | 0.87 | 0.52 | 0.45 | Bohai Sea | 14.4% | 9500 | 0.29 | 5640 |
| Dalian | 30% | 32% | 91% | 0.74 | 0.49 | 0.37 | | | | | |
| Lianyungang | 11% | 27% | 41% | 0.65 | 0.48 | 0.31 | | | | | |
| Qingdao | 91% | 99% | 92% | 3.01 | 0.47 | 1.41 | Yellow Sea | 67.3% | 5680 | 2.25 | -890 |
| Yantai | 100% | 100% | 100% | 4.17 | 0.54 | 2.24 | | | | | |

| City | | | | | | Sea | | | | |
|---|---|---|---|---|---|---|---|---|---|---|
| Shanghai | 80% | 100% | 80% | 2.23 | 0.45 | 1.00 | | | | |
| Ningbo | 56% | 58% | 95% | 3.51 | 0.48 | 1.7 | | | | |
| Wenzhou | 26% | 46% | 57% | 1.27 | 0.95 | 1.2 | East ChinaSea | 62.1% | 460 | 1.18 | -18170 |
| Fuzhou | 74% | 89% | 84% | 2.39 | 0.46 | 1.10 | | | | |
| Xiamen | 75% | 100% | 75% | 1.88 | 0.47 | 0.88 | | | | |
| Shenzhen | 100% | 100% | 100% | 1.00 | 0.90 | 0.90 | | | | |
| Zhuhai | 25% | 65% | 38% | 2.63 | 0.46 | 1.20 | | | | |
| Shantou | 24% | 100% | 24% | 1.30 | 0.88 | 1.14 | South ChinaSea | 43.4% | 9590 | 1.32 | -9060 |
| Zhanjiang | 28% | 52% | 53% | 2.49 | 0.48 | 1.19 | | | | |
| Beihai | 41% | 100% | 41% | 11.41 | 0.60 | 6.8 | | | | |
| Average | 48% | 70% | 64% | 2.48 | 0.57 | 1.37 | | | | |

For effectiveness verification of the evaluation, the average environmental performance of the cities which classified by their adjoining sea area was calculated ( Column I and K in Tab. 3) and compared with the actual observation of increased sea area with water quality below Grade 1(Column J and L in Tab. 3). It is found that they are matched quite well. In 2012 the average static environmental performance of the Yellow Sea and the East China Sea is higher than Bohai and the South China Sea's, and correspondingly, the sea area with water quality below Grade 1in Yellow Sea and East China Sea was increased smaller than in Bohai and South China Sea's from 2011 to 2012. Further observation found: the average dynamic environmental performance of the cities adjoining the Yellow Sea, East China Sea and South China Sea from 2010 to 2012 (in column K of Tab.3) is large than 1. It shows that the coastal environmental performance of these regions was improved. In the fact, from the statistics data of Tab.2, we found the sea area with water quality below Grade 1adjoining these cities was reduced from 2010 to 2012 (column L in Tab.3). but the average dynamic environmental performance of the cities adjoining Bohai is less than 1, and the sea area with water quality below Grade 1 in Bohai was increased.

The consistency of evaluation and observation shows that the index designed in this paper can better reflect the effect to the marine environment by economic construction in the cities. The index of coastal environmental performance will be better than other pure economic indicators to control pollution of the marine environment and promote economic development at the same time. On the other hand, the consistency also shows: the impact on the water quality of coastal waters is obvious for the industrial emissions generated by the coastal cities.

## 4. Results and discussion

For controlling pollution of the marine environment while developing coastal economy, the coastal environmental performance was proposed and measured in static and dynamic methods combined with DEA and efficiency theory in this paper. With the two methods, 16 harbor cities were evaluated. The results showed the index designed in this paper can better reflect the effect to the marine environment for economy of the coastal cities.

Further analysis calculation results in column B, C and D of Table 3, the static environmental performance and Static scale efficiency in 2012 are the highest in Yantai and Shenzhen, and static technical efficiency in 2012 (Column C in Tab.3) is higher in Qingdao, Yantai, Shanghai, Xiamen, Shenzhen and Shantou. Compared these values with Gross Regional Product (GRP)of these cities, it is found that the static scale efficiency tends to be relevant to GRP. The scale efficiency is often higher in the coastal cities which have higher GRP, such as Qingdao, Shenzhen; otherwise it is lower in the lower GRP cities, such as Zhuhai, Shantou. Because high scale efficiency is profitable for improving the static environmental performance, so it is helpful to promote the economic agglomeration for building a coastal economic center city. On the other hand, a smaller city may also increase its technical efficiency for improving the static environmental performance, such as Xiamen.

For the dynamic analysis of coastal environmental performance in column E, F and G of Table 3, Dynamic efficiency change of Tianjin, Qinhuangdao, Dalian, Lianyungang etc. decreases during this two years, while the other observation cities has increased. But further analyzing showed the improved efficiency change mainly depends on the dynamic efficiency change rather than the dynamic technical change. The average of the technical change is only 0.57, far less than 1. It indicates that improving the environmental technology is the most important task for further improvement of the environmental performance.